\documentclass[12pt]{article}
\usepackage{epsfig}\parskip 5pt plus 1pt
\newcommand{\be}{\begin{equation}}
\newcommand{\ee}{\end{equation}}
\newcommand{\bea}{\begin{eqnarray}}
\newcommand{\eea}{\end{eqnarray}}

\def\simge{\mathrel{%
   \rlap{\raise 0.511ex \hbox{$>$}}{\lower 0.511ex \hbox{$\sim$}}}}
\def\simle{\mathrel{
   \rlap{\raise 0.511ex \hbox{$<$}}{\lower 0.511ex \hbox{$\sim$}}}}
\newcommand{\helio}{^6{\rm He}}
\newcommand{\neon}{^{18}{\rm Ne}}

\begin{document}
\thispagestyle{empty}
\vspace*{1cm}
\begin{center}
{\Large{\bf A Beta Beam complex based on the machine
upgrades for the LHC } }\\
\vspace{.5cm}

A. Donini$^{\rm a}$, E. Fernandez-Martinez$^{\rm a}$,
P. Migliozzi$^{\rm b}$, S. Rigolin$^{\rm a}$, \\ L. Scotto Lavina$^{\rm
c}$, T. Tabarelli de Fatis$^{\rm d}$, F. Terranova$^{\rm e}$

\vspace*{1cm}
$^{\rm a}$ I.F.T. and Dep. F\'{\i}sica Te\'{o}rica, U.A.M., Madrid, Spain \\
$^{\rm b}$ I.N.F.N., Sezione di Napoli, Naples, Italy \\
$^{\rm c}$ Dip. di Fisica, Universit\`{a} "Federico II" and INFN,
Napoli, Italy \\
$^{\rm d}$  Universita di Milano Bicocca and I.N.F.N., Milano,
Italy \\
$^{\rm e}$ I.N.F.N., Laboratori Nazionali di Frascati,
Frascati (Rome), Italy \\

\end{center}

\vspace{.3cm}
\begin{abstract}
\noindent
The Beta Beam CERN design is based on the present LHC injection
complex and its physics reach is mainly limited by the maximum
rigidity of the SPS. In fact, some of the scenarios for the machine
upgrades of the LHC, particularly the construction of a fast cycling
1~TeV injector (``Super-SPS''), are very synergic with the
construction of a higher $\gamma$ Beta Beam.  At the energies that can
be reached by this machine, we demonstrate that dense calorimeters can
already be used for the detection of $\nu$ at the far location. Even
at moderate masses (40 kton) as the ones imposed by the use of
existing underground halls at Gran Sasso, the CP reach is very large
for any value of $\theta_{13}$ that would provide evidence of $\nu_e$
appearance at T2K or NO$\nu$A ($\theta_{13}\geq 3^\circ$). Exploitation of matter effects at
the CERN to Gran Sasso distance provides sensitivity to the neutrino
mass hierarchy in significant areas of the $\theta_{13}-\delta$ plane.
\end{abstract}

\vspace*{\stretch{2}}
\begin{flushleft}
  \vskip 2cm
{ PACS: 14.60.Pq, 14.60.Lm}
\end{flushleft}

\newpage

\section{Introduction}
\label{introduction}
Since a few years, we have solid experimental
evidence~\cite{ref:meas_alpha_sol,ref:meas_alpha_atm} that the ratio
$\Delta m^2_{12}/|\Delta m^2_{23}|$ of the neutrino squared mass
differences driving the solar and atmospheric oscillations is of the
order of $ {\cal O}(10^{-2})$. This measurement has an enormous impact
in the design of future experiments posed to thoroughly determine the
leptonic mixing matrix (PMNS~\cite{PMNS}). In particular, given a
relatively large $\Delta m^2_{12}/|\Delta m^2_{23}|$ ratio, the
determination of the currently unknown 1-3 sector of the PMNS, i.e.
the mixing between the first and third generation and the CP violating
Dirac phase,
can be accomplished by long baseline experiments measuring the
contamination of $\nu_\mu \rightarrow \nu_e$
transitions in the bulk of $\nu_\mu \rightarrow \nu_\tau$ oscillations
at the atmospheric scale. The size of these sub-dominant contributions
depends on the mixing angle between the first and third neutrino
generation ($\theta_{13}$) and an experimental determination of this
angle is mandatory to establish to what extent future facilities are
able to address CP violation in the leptonic sector or fix the
neutrino mass hierarchy (sign of $\Delta m^2_{23}$) exploiting matter
effects. Should the size of $\theta_{13}$ be large enough to allow the
observation of $\nu_\mu \rightarrow \nu_e$ oscillations at the
atmospheric scale in the forthcoming
experiments~\cite{ref:future_exp,T2K,nova} ($\theta_{13} \simge
3^\circ$), new facilities would be needed to close up the PMNS. They
should perform precision measurements of the 1-3 sector and
particularly of the CP violating phase. In this context a novel
neutrino source like the Beta Beam~\cite{Zucchelli:sa} (BB) offers
unprecedented opportunities thanks to the large intensities and
purities available. Moreover, it represents a unique European facility
since it could leverage the present CERN acceleration
complex. Unfortunately, in its present design~\cite{Bouchez:2003fy}
the physics potential of the Beta Beam is not fully
exploited~\cite{Burguet-Castell:2003vv}. The maximum rigidity of the
CERN SPS machine limits the energy of the outgoing neutrinos, so that
very large detectors are needed to overcome the smallness of the cross
sections. Tuning the oscillation probability to the first peak, the corresponding baseline is too short to exploit matter
effects and, hence, determine the sign of $\Delta m^2_{23}$. Moreover,
dense detectors cannot be employed to separate $\nu_\mu$ from $\nu_e$
interactions, so that enormous underground facilities must be built on
purpose.

On the other hand, it is unlikely that the CERN acceleration complex
will remain unchanged up to the time of operation considered in the
baseline design (about 2020~\cite{ref:bouchet_nnn}). In fact, some of
the scenarios for the machine upgrades of the LHC are, accidentally,
very synergic with a higher energy Beta Beam. In this paper, we
identify the options that could leverage a strong neutrino programme
aimed at a full determination of the PMNS and $\nu$ mass hierarchy, and
the setups that can exploit existing underground facilities and
moderately massive dense detectors (Sec.\ref{sec:machine},
\ref{sec:rates} and \ref{sec:detector}). For any value of
$\theta_{13}$ that allows evidence for $\nu_\mu \rightarrow \nu_e$
oscillations in the forthcoming experiments up to T2K~\cite{T2K} or
NO$\nu$A~\cite{nova}, we show that these setups have the sensitivity to
address CP violation in the leptonic sector
(Sec.\ref{sec:sensitivity}). We also compute the minimum intensity
required to guarantee full coverage of the T2K sensitivity region. We
draw our conclusions in Sec.~\ref{sec:conclusions}.

\section{The accelerator complex}
\label{sec:machine}

The steady progress in the technology of radioactive ion production
and acceleration opens up the possibility of obtaining pure sources of
$\nu_e$ directly from $\beta$ unstable
isotopes~\cite{Zucchelli:sa}. These sources have practically no
contamination from other flavors and a well defined energy spectrum
that depends on the kinematics of $\beta$ decay. The choice of the isotope
is a compromise between production yield, $Q$ value and
lifetime. Isotopes with short lifetimes, $\tau \simle $1s, must be
handled by fast cycling boosters to avoid strong losses during the
acceleration phase. On the other hand, larger lifetimes result in a
decrease of the number of decays per unit time (hence, of $\nu$ flux)
for a fixed number of ion stacked. The best isotopes identified so far
are $\helio$ for antineutrino production (a $\beta^-$ emitter with
$E_0=3506.7$~keV and a 806.7 ms half life) and $\neon$ for neutrinos
($E_0=3423.7$~keV and half life of 1.672~s). The former is obtained by
neutron absorption in beryllium oxide and requires a
proton-to-neutron converter; the latter is produced through
spallation, e.g. from proton interactions with a magnesium oxide
target\footnote{Recently a novel method to produce high intensity ($^{8}$Li and $^{8}$B) beams has been put forward in Ref.~\cite{rubbione}. Using these beams interesting physics results can be obtained~\cite{cocktail}.}. Both require a $\sim 200$~kW proton driver operating in the
few GeV region. The collection and ionization of the ions is performed
using the ECR technique. Hereafter ions are bunched, accelerated and
injected up to the high energy boosters. In the baseline design, the
proton driver is the proposed Super Proton Linac (SPL)~\cite{SPL}. The
SPL is a multi-megawatt ($\sim 4$~MW, $E_p=$2.2~GeV~\cite{SPL} or
3.5~GeV~\cite{Garoby:2005se,Campagne:2004wt}) machine aimed at
substituting the present Linac2 and PS Booster (PSB). Contrary to
naive expectation, a multi-megawatt booster is not necessary for the
construction of a Beta Beam or a nuclear physics
(EURISOL-like~\cite{betabeams_moriond}) facility and could be fully
exploited only by a low-energy neutrino
SuperBeam~\cite{Mezzetto:2003mm} or by a Neutrino Factory complex.
Any of the possibilities currently under discussion at
CERN~\cite{dainton,garoby_hif04} for the upgrade of the PSB based
either on Rapid Cycling Syncrotrons or on Linacs represents a viable
solution for the production stage of a Beta Beam complex. They would
allow production of $\sim 2{\times} 10^{13}$ $\helio$/s for 200~kW on
target, consistently with the current SPL-based design.  A proper
upgrade of the sub-GeV boosters represents an important step toward
full exploitation of the LHC physics capabilities; in particular, the
luminosity of the Large Hadron Collider would highly benefit from a
modification of the pre-injectors, which are currently limited by
space charge at the PSB and PS injection energies~\cite{bruning}.  The
choices and timescale for the upgrades of the LHC will depend on the
feedbacks from the first years of data taking. Still, three phases can
already be envisaged~\cite{bruning,scandale}: an optimization of
present hardware (``phase 0'') to reach the ultimate luminosity of
$2{\times} 10^{34}$~cm$^{-2}$s$^{-1}$ at two interaction points; an
upgrade of the LHC insertions (``phase 1'') and, finally, a major
hardware modification (``phase 2'') to operate the LHC in the ${\cal
L} \simeq 10^{35}$~cm$^{-2}$s$^{-1}$ regime and, if needed, prepare
for an energy upgrade. The most straightforward approach to ``phase
2'' would be the equipment of the SPS with fast cycling
superconducting magnets in order to inject protons into the LHC with
energies of about 1~TeV. At fixed apertures at injection, this option
would double the peak luminosity of the LHC.  Moreover, injecting at
1~TeV strongly reduces the dynamic effects of persistent currents,
ease stable operation of the machine and, therefore, impact on the
integrated luminosity of the collider. The 1~TeV injection option
(``Super-SPS'') would have an enormous impact on the design of a Beta
Beam at CERN. This machine fulfills simultaneously the two most
relevant requirements for a high energy BB booster: it provides a fast
ramp ($dB/dt=1.2 \div 1.5$~T/s~\cite{fabbricatore}) to minimize the
number of decays during the acceleration phase and, as noted in the first
Reference of \cite{Burguet-Castell:2003vv}, it is able to bring $\helio$ up to
$\gamma \simeq 350$ ($\neon$ to $\gamma \simeq 580$). In this case,
neutrinos are produced by the Beta Beam with energies of the order of
few GeV ($\langle E \rangle =$ 2.18 GeV for neutrinos, 1.35 for
antineutrinos). As shown in Sec.\ref{sec:detector}, this energy allows
the use of dense detectors and existing underground infrastructures,
the first peak of oscillation probability being comparable to the CERN to
Gran Sasso distance. Due to the large increase of the cross-section, a
strong reduction of the detector mass is possible compared with the
baseline design. Clearly, the exploitation of the Super-SPS as a final
booster for the BB is not in conflict with LHC operations, since the
Super-SPS operates as injector only for a small fraction of its duty
time (LHC filling phase).

The use of the Super-SPS as the final booster of the BB is not the
only possibility that can be envisaged to reach the multi-GeV
regime. After injection of the ions from the SPS to the LHC, a
mini-ramp of the LHC itself would bring the ions at
$\gamma=350-580$. Differently from the previous case, however, this
option would require allocation of a significant fraction of the LHC
duty cycle for neutrino physics and could be in conflict with ordinary
collider operations.  Moreover, this option requires dedicated machine
studies to quantify the injection losses or optimize the dipole
ramp. For these reasons, in the following we mainly focus on the
exploitation of the Super-SPS\footnote{It is worth mentioning that the
Super-SPS eases substantially injection of $\beta$-unstable ions in
the LHC to reach $\gamma \gg 350-580$. For a discussion of this option
we refer to~\cite{Terranova:2004hu}.}.

\begin{figure}
\centering
\epsfig{file=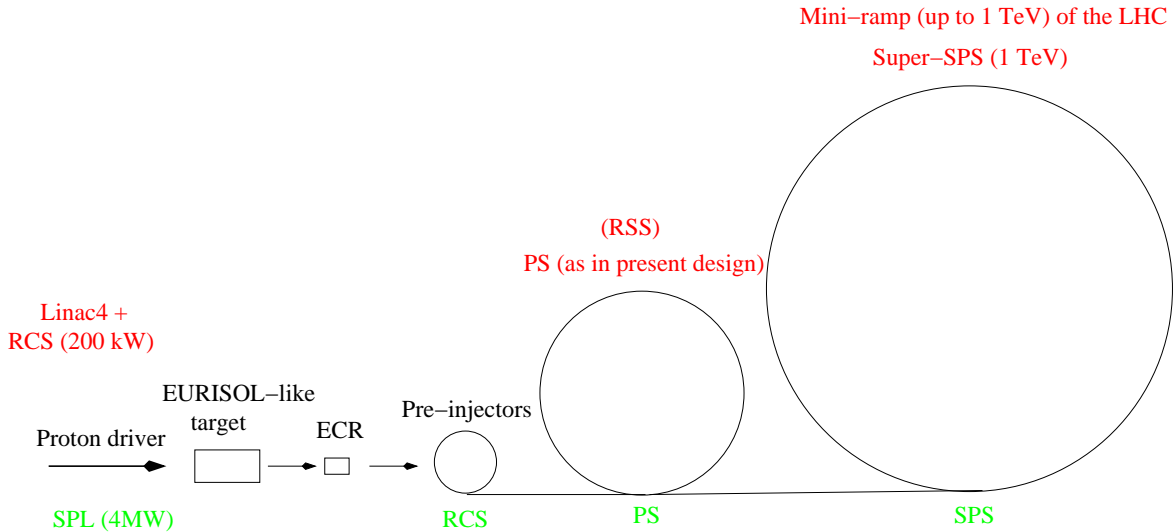,width=\textwidth}
\caption{The main components of the Beta Beam complex up to injection
into the decay ring. In the lower part, the machines considered in the
baseline option are indicated. The alternatives that profit of the
upgrade of the LHC injection system are also mentioned (upper part).
RCS stands for Rapid Cycling Syncrotron, RSS for Rapid Superconducting
Syncrotron~\cite{garoby_hif04}. Other abbreviations are defined in the
text.}
\label{fig:complex}
\end{figure}

The increase of the ion energy in the last element of the booster
chain represents a challenge for stacking~\cite{terranova_nufact04}.
Ions of high rigidity must be collected in a dedicated ring of
reasonable size. In the baseline design, this is achieved by a decay
ring made of small curved sections (radius $R\sim 300$~m) followed by
long straight sections ($L=2500$~m) pointing toward the far neutrino
detector.  In this case, the decays that provide useful neutrinos are
the ones occurring in the straight secsion where neutrinos fly in the
direction of the detector and the useful fraction of decays
(``livetime'') is limited by the decays in the opposite arm of the
tunnel. For the CERN to Frejus design the livetime is $L/(2\pi R+2L)
\sim 36$\% and the overall length has been fixed to 6880~m. A decay
ring of the same length equipped with LHC dipolar magnets (8.3~T)
would stack ions at the nominal Super-SPS rigidity with a significantly
larger radius ($\sim 600$~m). The corresponding livetime is thus
23\%. Again, current R\&D related with the LHC upgrades and aimed at
the development of high field magnets
(11$\div$15~T)~\cite{bruning,Devred:2005vd} can be used to reduce the costs for the decay ring and
increase the livetime \footnote{The correspondence
between the various magnet R\&D and the BB is not accidental: in a BB
complex the booster plays the role of the collider injector and,
hence, profits of the requirements for fast cycling; the decay ring
plays the role of the collider, which is aimed at the highest possible
rigidity, even at the expense of the ramp time. Clearly, ramping speed
is immaterial for the BB stacking ring.}.  An additional reduction of
flux comes from the increase of ion lifetime (about a factor of three
for $\helio$ at $\gamma=350$ compared with the baseline option at
$\gamma=100$~\cite{lindroos_nnn}). There is, however, a strong
correlation between the increase of the neutrino energy and the amount
of ions that can be stacked into the decay ring. In the baseline
design, the constraint on the number of circulating bunches and on
the bunch length comes from the need of timing the parent ion. This is
mandatory to suppress the atmospheric background in the far
detector. The smaller the time occupancy of the ion bunches in the
ring, the larger the suppression factor (SF):
\be
\mathrm{SF} \ = \ \frac{\Delta t_b {\cdot} N_b {\cdot} v}{2\pi R+2L}
\ee
$\Delta t_b$ being the time length of the bunch, $N_b$ the number of
circulating bunches, $v\simeq c$ the ion velocity and $2\pi R+2L$ the
length of the ring. In the baseline design this value must be kept at
the level of $10^{-3}$, implying a challenging $\Delta t_b$=10~ns time
structure of the bunch for $N_b=8$ circulating bunches. At higher
energies (e.g. $\gamma=350$ for $\helio$) the atmospheric background
is suppressed by about one order of magnitude and the SF can be
correspondingly relaxed, provided that the injection system can
sustain the increased request of bunches and/or ions per bunch. Since a
complete machine study concerning this issue is still
missing\footnote{For recent progresses in the framework of the
baseline design (SPS-based), see~\cite{nota_lindroos}.} - especially
for the Super-SPS - in the following, physics performances are
determined as a function of fluxes.  We remind that the baseline BB
design aims at $2.9{\times} 10^{18}~\helio$ and $1.1{\times}
10^{18}~\neon$ decays per year (``nominal intensity'').
Fig.\ref{fig:complex} sketches the main components of the BB complex
up to injection into the decay ring. In the lower part, the machines
considered in the baseline option are listed. The alternatives that
profit of the upgrade of the LHC injection system are also mentioned
(upper part).

\section{Far detector concept and expected rates}
\label{sec:rates}

Traditional technologies for $\nu$ production (up to the so-called
``Superbeams'') allow the investigation of the 1-3 sector of the
leptonic mixing matrix through the appearance of $\nu_e$ and
$\bar{\nu}_e$ at baselines $\ge 100$~km, i.e. through the information
coded in the $\nu_\mu \rightarrow \nu_e$ and $\bar{\nu}_\mu
\rightarrow \bar{\nu}_e$ transitions probabilities. In this context,
optimal far detectors are low-density, massive e.m. calorimeters
(liquid scintillators, water Cherenkov, liquid Argon
TPC's~\cite{nostrareview}). On the other hand, both the Beta Beams and
the Neutrino Factories~\cite{neutrinofactories} exploit the
T-conjugate channel $\nu_e \rightarrow \nu_\mu$ and $\bar{\nu}_e
\rightarrow \bar{\nu}_\mu$.  In the case of the Beta Beam, oscillated
neutrinos are the {\it only} source of primary muons while for the
neutrino factory the detector must be able to identify the muon charge
with outstanding efficiencies ($>99.9$\%) to distinguish the $\nu_e
\rightarrow \nu_\mu$ signal from the $\bar{\nu}_\mu$ background. In
both cases, calorimetric measurements are needed to reconstruct the
neutrino energy\footnote{The only notable exception concerns the
``monochromatic Beta Beams''\cite{monocromatic} based on ions decaying
through electron capture.}. In particular, the choice of the passive
material of the calorimeter depends on the typical range of the
primary muon; the latter must be significantly larger than the
interaction length to allow for filtering of the hadronic part and
effective NC and $\nu_e$~CC selection. For neutrinos of energies
greater than $\sim$1~GeV, iron offers the desired properties. As a
consequence, the energy of the Super-SPS can be exploited to switch
from a low-Z to a high-Z/high-density calorimeter also in the case of
the Beta Beam.  The use of iron detectors avoids the need for large
underground excavations, which are mandatory for Beta Beams at lower
$\nu$ energies. Since these detectors are capable of calorimetric
measurements, they can be exploited even better than water Cherenkov to
obtain spectral informations. They're not expected to reach, anyhow,
the granularity of liquid argon TPC's or the megaton-scale mass of
water Cherenkov's; hence, in spite of the underground location, they
cannot be used for proton decay measurements and low-energy
astroparticle physics.

Magnetization of the iron is not strictly necessary,
even if it contributes to reducing the pion punch-through background (the
sign of the primary muon is uniquely determined by the ion species
circulating in the stacking ring). Since the neutrino energy at the
Super-SPS matches the CERN to Gran Sasso baseline, a high density
calorimeter can be hosted into the existing halls of the Gran Sasso
laboratories up to fiducial masses of $\sim$40~kton.

%
\begin{figure}[t!]
\vspace{-0.5cm}
\begin{center}
\begin{tabular}{c}
\hspace{-0.3cm} \epsfxsize10cm\epsffile{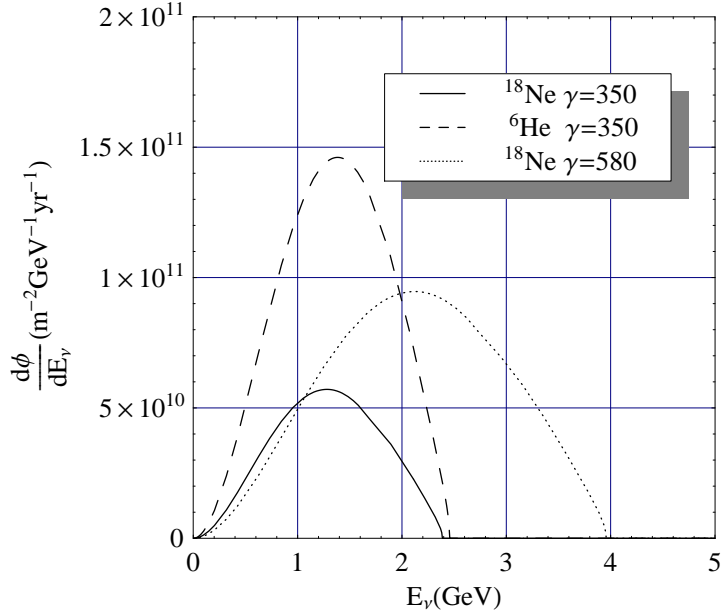}\\
\end{tabular}
\caption{\label{fig:fluxes} Beta Beam fluxes at the Gran Sasso
location (735 km baseline) as a function of the neutrino energy.}
\end{center}
\end{figure}


Fig.\ref{fig:fluxes} shows the corresponding neutrino fluxes (nominal
intensity) at the Gran Sasso location for both $\helio$ and $\neon$ at
$\gamma = 350$ (``$\gamma=350,350$ option'') and for $\neon$ at the
Super-SPS maximum rigidity ($\gamma=580$: ``$\gamma=350,580$
option'').  The calculation include the effects of finite electron
mass and the three different decay modes of $^{18}$Ne, each with a
different end-point energy, see Table~\ref{betabeam}.

\begin{table}
\begin{center}
\begin{tabular}{|c|c|c|} \hline \hline
   Element & End-Point (MeV) & Decay Fraction \\ \hline
            & 34.114 & 92.1\% \\
  $^{18}$Ne & 23.699 & 7.7\% \\
            & 17.106 & 0.2\% \\ \hline
 $^{6}$He   & 35.078 & 100\% \\ \hline
\hline
\end{tabular}
\caption{\label{betabeam} $^{18}$Ne and $^{6}$He $\beta$-decay channels and relative
end-point energies from~\cite{betadecays}.}
\end{center}
\end{table}

%
%


A relevant source of uncertainty in the determination of the event
rates is the present poor knowledge of the $\nu N$ and $\bar \nu N$
cross-sections for energies below 1~GeV~\cite{Zeller:2003ey}: either
there are very few data (the case of neutrinos) or there are no data
at all (the case of antineutrinos).  On top of that, the few available
data have generally not been taken on the target used in the
experiments (either water, iron, lead or plastics), and the
extrapolation from different nuclei is non trivial. The situation
improves at the energies relevant for the Super-SPS although this
region is still below the DIS dominated regime. Note, moreover, that a
much improved knowledge of the cross sections will be available in the
forthcoming years both due to a dedicated experimental
campaigns~\cite{Zeller_nufact05} and from the near detectors of the
next generation long baseline experiments.

We use in this paper the cross-section on iron obtained following Ref.~\cite{lipari}. In table \ref{table:rates} the expected unoscillated $\nu_e$ CC events per kton-year are shown for $\gamma=350,580$ together with the fraction of QE, RES and DIS in the sample.

\begin{table}
\begin{center}
\begin{tabular}{l|r|r|r} \hline 
       &  $\nu$ (350) & $\nu$ (580) &  $\bar{\nu}$ (350) \\
\hline
DIS    &  1.57   &  75.18   &  0.06   \\
RES    &  16.76  &  60.82   &  29.72  \\
QE     &  37.47  &  115.37  &  30.28  \\
Total  &  55.80  &  251.36  &  60.07  \\
\hline
\end{tabular}
\caption{Expected unoscillated $\nu_e$ CC events per kton-year.}
\end{center}
\label{table:rates}
\end{table}

\section{The iron calorimeter}
\label{sec:detector}

Several techniques can be employed for the design of the active
detectors of large mass iron calorimeters. Due to cost constraints,
most of the options are based on plastic scintillators or gaseous
detectors.  In the present study, we consider a design derived from a
digital hadron calorimeter proposed for the reconstruction of the
energy flow at the ILC detector~\cite{DHCAL1,tdf} and based on glass
Resistive Plate Chambers (DHCAL). In this case, gas detectors are
particularly appealing since they allow highly granular designs. On
the other hand, in the context of the Beta Beam the advantages mainly
reside on the low production cost of RPC, along the line investigated
by the MONOLITH~\cite{monolith} and INO~\cite{ino}
collaborations. Clearly, a systematic comparison of the options
available is beyond the scope of the present work.

\begin{figure*}[tbph]
\centering
\includegraphics[width=75mm]{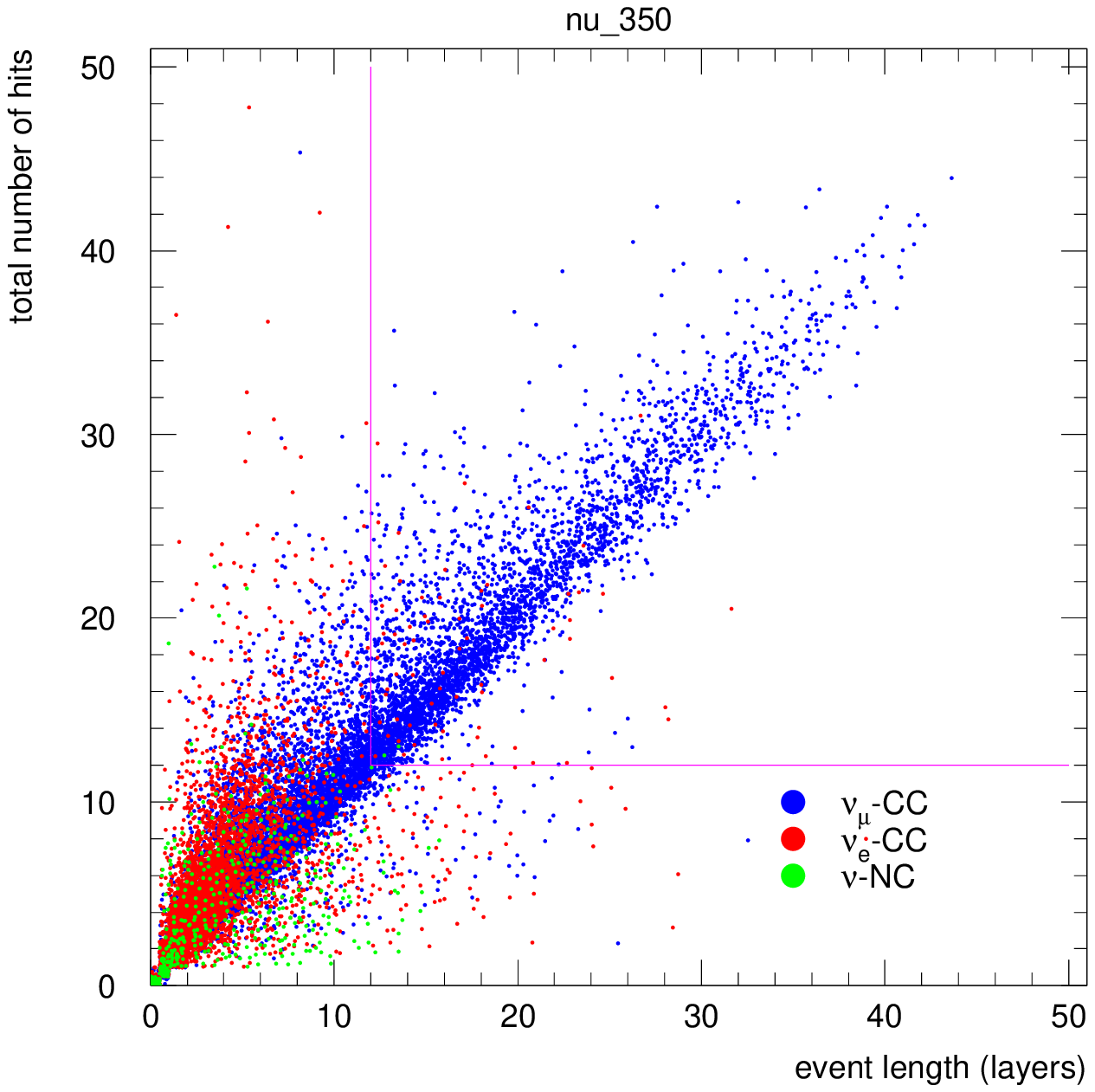}\includegraphics[width=75mm]{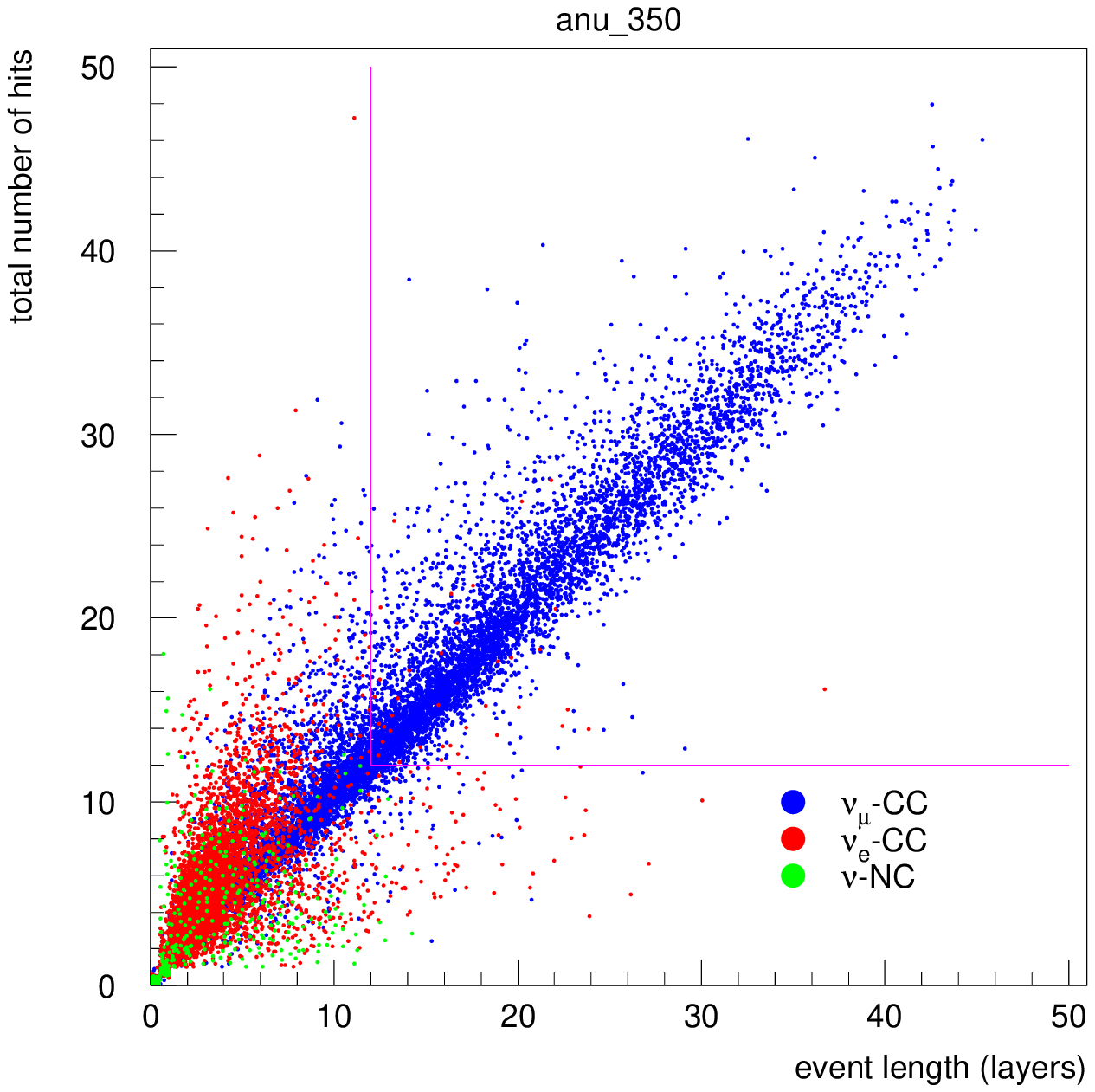}
\caption{Scatter plot of the total number of hits recorded in the detector versus
the total length (given in number of crossed layers) of the event for
neutrinos (left) and anti-neutrinos (right) with $\gamma=350$.}
  \label{fig:eveclass}
\end{figure*}

The configuration considered hereafter consists of a sandwich of 4~cm
non-magnetized iron interleaved with glass RPC's to reach an overall
mass of 40~kton. The RPC are housed in a 2~cm gap; the active element
is a 2~mm gas-filled gap; the drift field is produced by 2 mm thick
glass electrodes coated with high resistivity graphite.  The signal is
readout on external pick-up electrodes segmented in $2 \times 2$
~cm$^2$ pads, providing a single-bit information. A full
GEANT3~\cite{GEANT3} simulation of this geometry has been implemented
along the lines discussed in~\cite{tdf}, including a coarse
description of the RPC materials and an approximate description of the
digitization process.  Spark generation in the active medium is
assumed to happen with 95\% efficiency upon the passage of an ionising
particle. As detailed in~\cite{tdf}, the accuracy of this simulation
has been validated by comparing its predictions to existing data
collected with a small prototype exposed to a pion beam of energy from
2 GeV to 10 GeV~\cite{gustavino}.  Inclusive variables (total number
of hits and event length expressed in terms of number of crossed iron
layers) have been used for event classification\footnote{The event
discrimination capability can be further improved developing a
dedicated pattern recognition system aimed at identifying explicitly
hits belonging to the primary muons.}. The scatter plot of the event
length versus the total number of hits of the event is shown in
Fig.~\ref{fig:eveclass} for neutrinos (left panel) and anti-neutrinos
(right panel) both coming from ions accelerated at $\gamma= 350$.
$\nu_\mu$ and $\nu_e$ charged-current (CC) interactions as well as
neutrino neutral-current (NC) interactions are shown with different
colors. An interaction is classified as a $\nu_\mu$ CC-like event if
both the event length and the total number of hits in the detector are
larger than 12. In the case the $\neon$ is ran at $\gamma= 580$, we
classify an event as a CC-like interaction if the event length and the
total number of hits are larger than 15 and 17, respectively. The
typical efficiency for identifying a neutrino or anti-neutrino CC
interaction averaged out over the whole spectrum is of the order of
50-60\%. Conversely, the probability for the background to be
identified as a CC-like event is slightly less than 1\%. Compared with
more challenging beta beams or neutrino factory designs, this facility
offers a limited pion rejection capability; this is due to the low energy
of the neutrinos (compared with the neutrino factory or very high
gamma Beta Beam options) and the choice of a dense detector. Still, as
we demonstrate in the subsequent sections, these performances are
sufficient to explore CP violation in the leptonic sector for any
value of $\theta_{13}$ that gives a positive $\nu_e$ appearance signal
in the next generation of long-baseline experiments: MINOS, OPERA and
the so-called ``Phase I'' experiments based on Superbeams (T2K, NO$\nu$A)
or reactors.

\begin{figure*}[tbph]
\centering
\includegraphics[width=75mm]{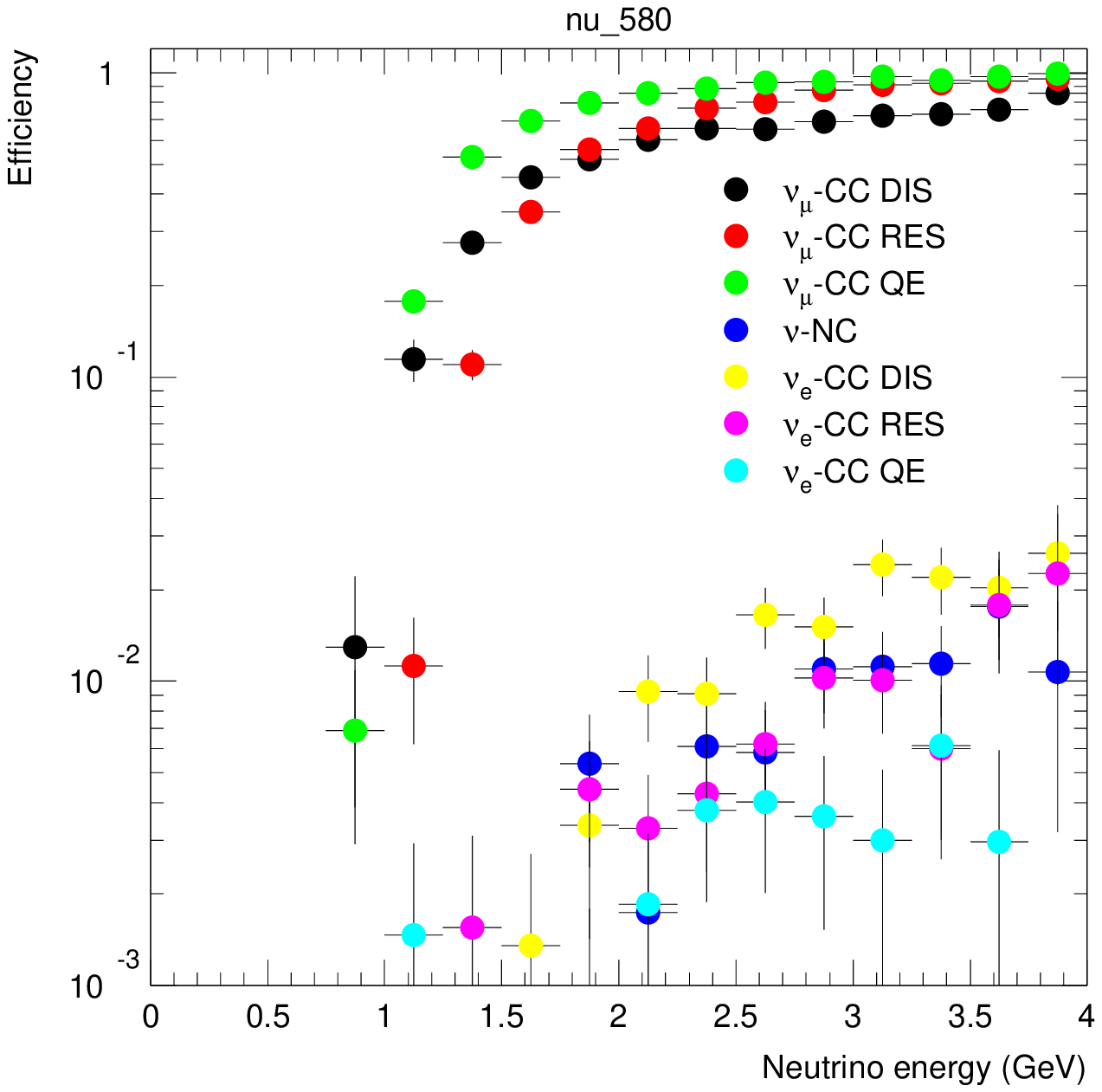}\includegraphics[width=75mm]{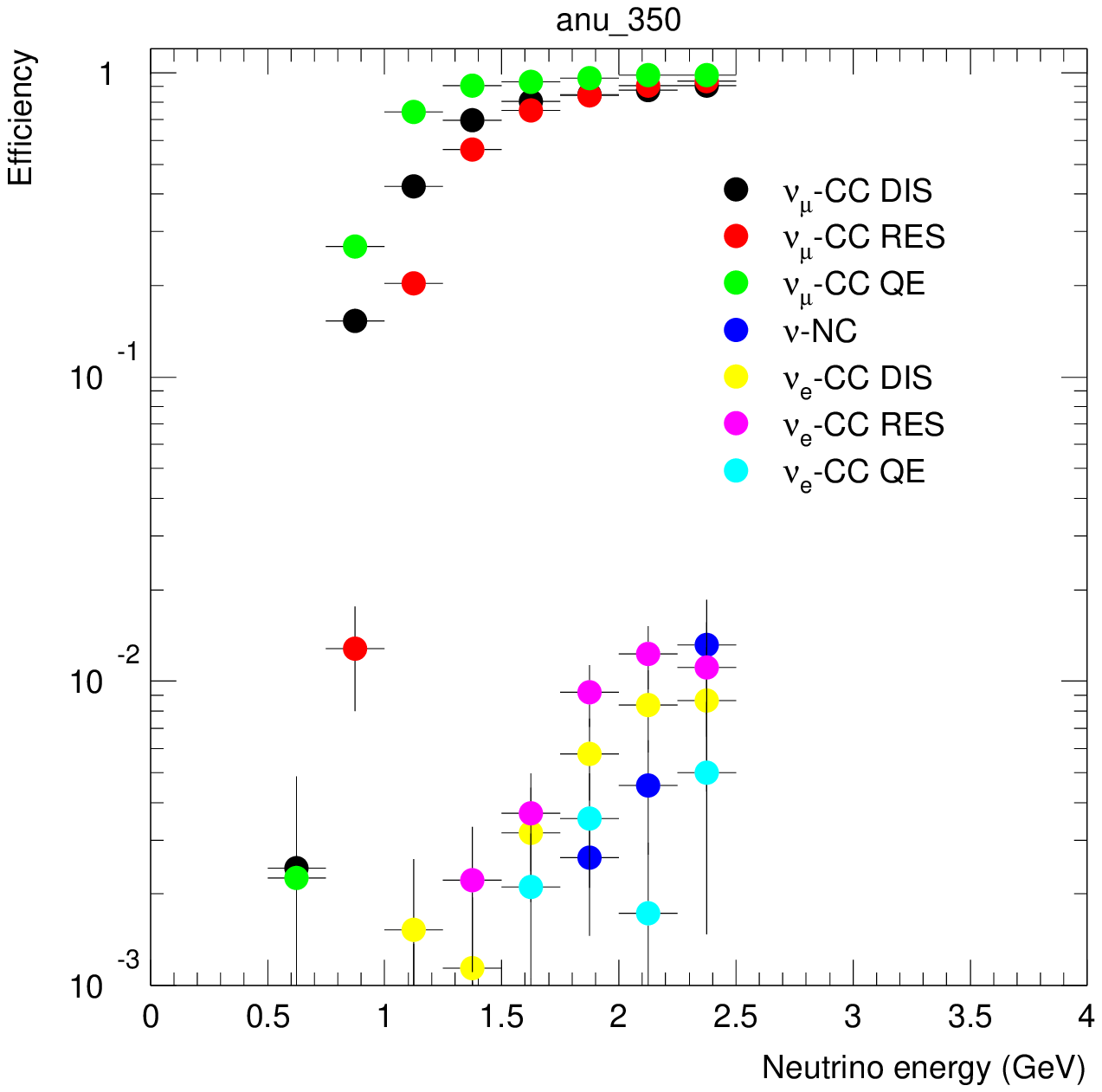}
\caption{Efficiencies for the signal ($\nu_\mu$ and $\bar{\nu}_\mu$
charged-current interactions) to be identified as CC-like event and
for the background ($\nu_e$ and $\bar{\nu}_e$ interactions, and
$\nu_\mu$ and $\bar{\nu}_\mu$ neutral-current interactions) to be
mis-identified as a CC-like events.}
  \label{fig:eveclasseff}
\end{figure*}

Finally, the efficiencies to correctly identify $\nu_\mu$
and $\bar{\nu}_\mu$ charge-current interactions are shown
for deep-inelastic (DIS), quasi-elastic (QE) and resonance (RES)
production, separately, in Fig.~\ref{fig:eveclasseff} as well as the probability
that $\nu_e$ and $\bar{\nu}_e$, separately for DIS, QE and RES
production, and neutral-current interactions are identified as a
CC-like interaction.

\section{Evaluation of the physics reach}
\label{sec:sensitivity}

The currently unknown mixing parameters $\theta_{13}$ and $\delta$
determine the inclusive rate of $\nu_\mu$ ($N^-$) and $\bar{\nu}_\mu$
($N^+$) CC events at the far location. Matter effects, moreover, are
sizable at baselines comparable to the CERN to Gran Sasso distance and
introduce additional modifications.  In particular, for positive
(negative) mass hierarchy and positive (negative) values of the Dirac
CP phase, it is possible to determine the sign of $\Delta m^2_{23}$
from the simultaneous measurements of $N^+$ and $N^-$. Matter effects
induce also spectral distortions in the $\nu_\mu$ and $\bar{\nu}_\mu$
distributions that improve the sensitivity and help in the ambiguous
region of negative (positive) CP phases. In the following,
sensitivities are computed from a binned likelihood fit of the $N^+$
and $N^-$ samples. The likelihood incorporates the finite energy
resolution of the detector (migration matrices) computed from the full
simulation of Sec.\ref{sec:detector}. Atmospheric background is
neglected together with the systematics affecting the $\nu /
\bar{\nu}$ ratio and a 2\% systematic uncertainty in the detector
efficiency is assumed.

The expected number of events as a function of $\delta$ and
$\theta_{13}$ are shown in Table~\ref{event} for nominal fluxes. For
the already measured oscillation parameters, we assumed the following
values: $\Delta m^2_{12} = 8.2{\times}10^{-5}~\mbox{eV}^2;\,
\theta_{12} = 33^\circ;\, \Delta m^2_{23} =
2.5{\times}10^{-3}~\mbox{eV}^2$ \cite{Gonzalez-Garcia:2004jd}. In the rest of this paper we only consider positive $\Delta m^2_{23}$ and $\theta_{23} = 45^\circ$, in this way only the intrinsic degeneracy is accounted for.

\begin{table}[hbt]
\begin{center}

\begin{tabular}{l|r|r|r|r|r|r|r}
                                                     
\hline
$\theta_{13}$ & $\delta$  & $\nu_\mu$CC &
$\nu_\mu$CC & $\bar{\nu}_\mu$CC & $\nu$-back. &
$\nu$-back. & $\bar{\nu}$-back.\\
&  &  $\gamma=350$ & $\gamma=580$ &  & $\gamma=350$ &  $\gamma=580$
& \\
                                                               
\hline
$  1^\circ$ & $-90^\circ$ &    1.43 &    9.44 &   37.33 &  126.02 &
881.11 &   77.28 \\
$  5^\circ$ & $-90^\circ$ &  105.44 &  485.48 &  266.89 &  126.02 &
881.11 &   77.28 \\
$ 10^\circ$ & $-90^\circ$ &  541.17 & 2291.01 &  848.62 &  126.02 &
881.11 &   77.28 \\
$  1^\circ$ & $  0^\circ$ &   18.27 &   80.45 &   22.58 &  126.02 &
881.11 &   77.28 \\
$  5^\circ$ & $  0^\circ$ &  189.51 &  840.07 &  193.23 &  126.02 &
881.11 &   77.28 \\
$ 10^\circ$ & $  0^\circ$ &  708.67 & 2997.51 &  701.85 &  126.02 &
881.11 &   77.28 \\
$  1^\circ$ & $ 90^\circ$ &   32.23 &   99.40 &    2.61 &  126.02 &
881.11 &   77.28 \\
$  5^\circ$ & $ 90^\circ$ &  259.27 &  934.72 &   93.49 &  126.02 &
881.11 &   77.28 \\
$ 10^\circ$ & $ 90^\circ$ &  847.67 & 3186.08 &  503.13 &  126.02 &
881.11 &   77.28 \\
\hline

\end{tabular}
\label{event}
\end{center}
\caption{ Event rates for a 10 years exposure.  The observed
oscillated CC events for different values of $\delta$ and
$\theta_{13}$ are given assuming normal neutrino mass hierarchy and
$\theta_{23}=45^\circ$. The expected background is also reported.}
\end{table}

\subsection{Establishing CP violation in the leptonic sector}

\begin{figure*}[tbhp]
\centering
\includegraphics[width=75mm]{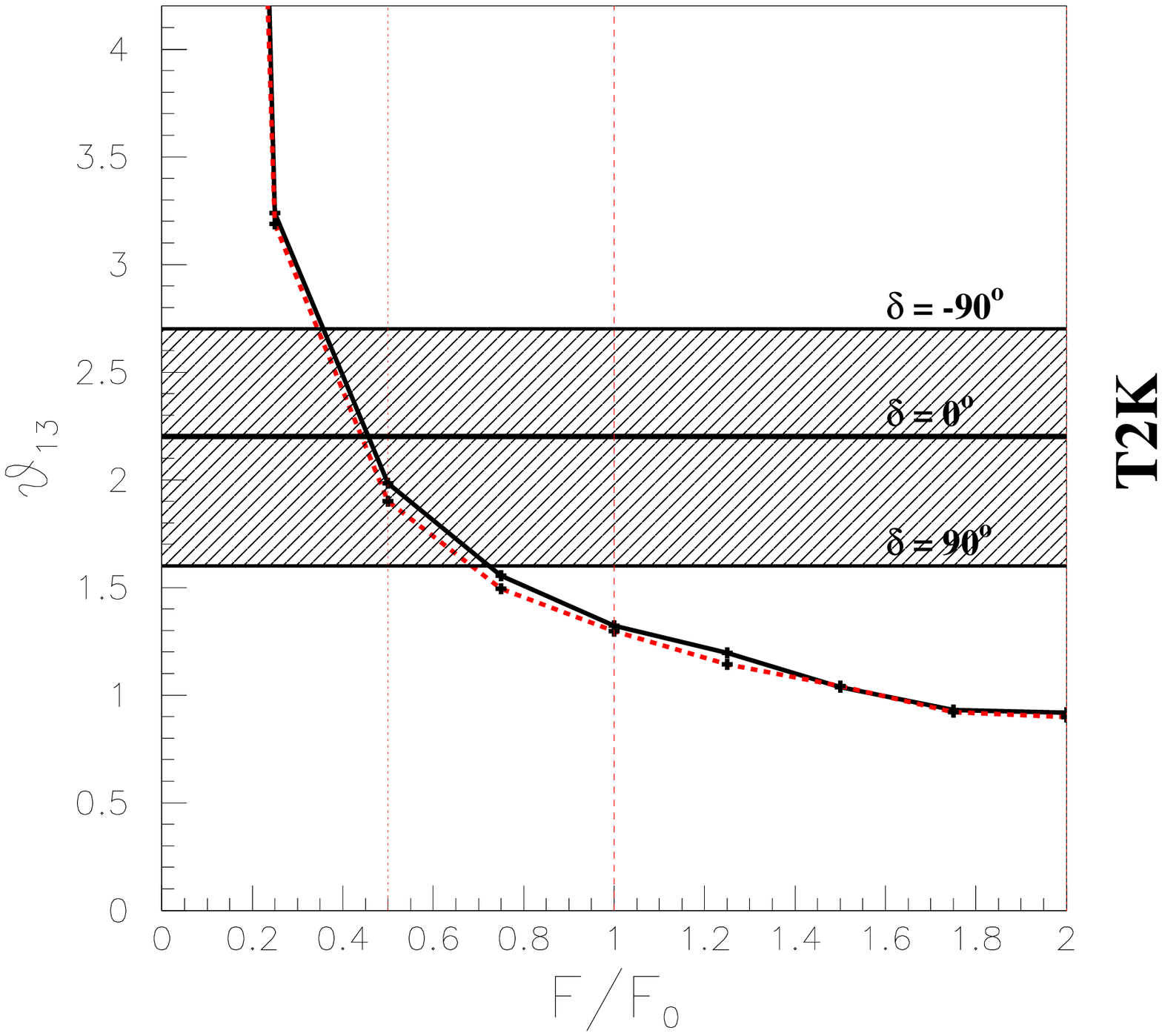}\includegraphics[width=75mm]{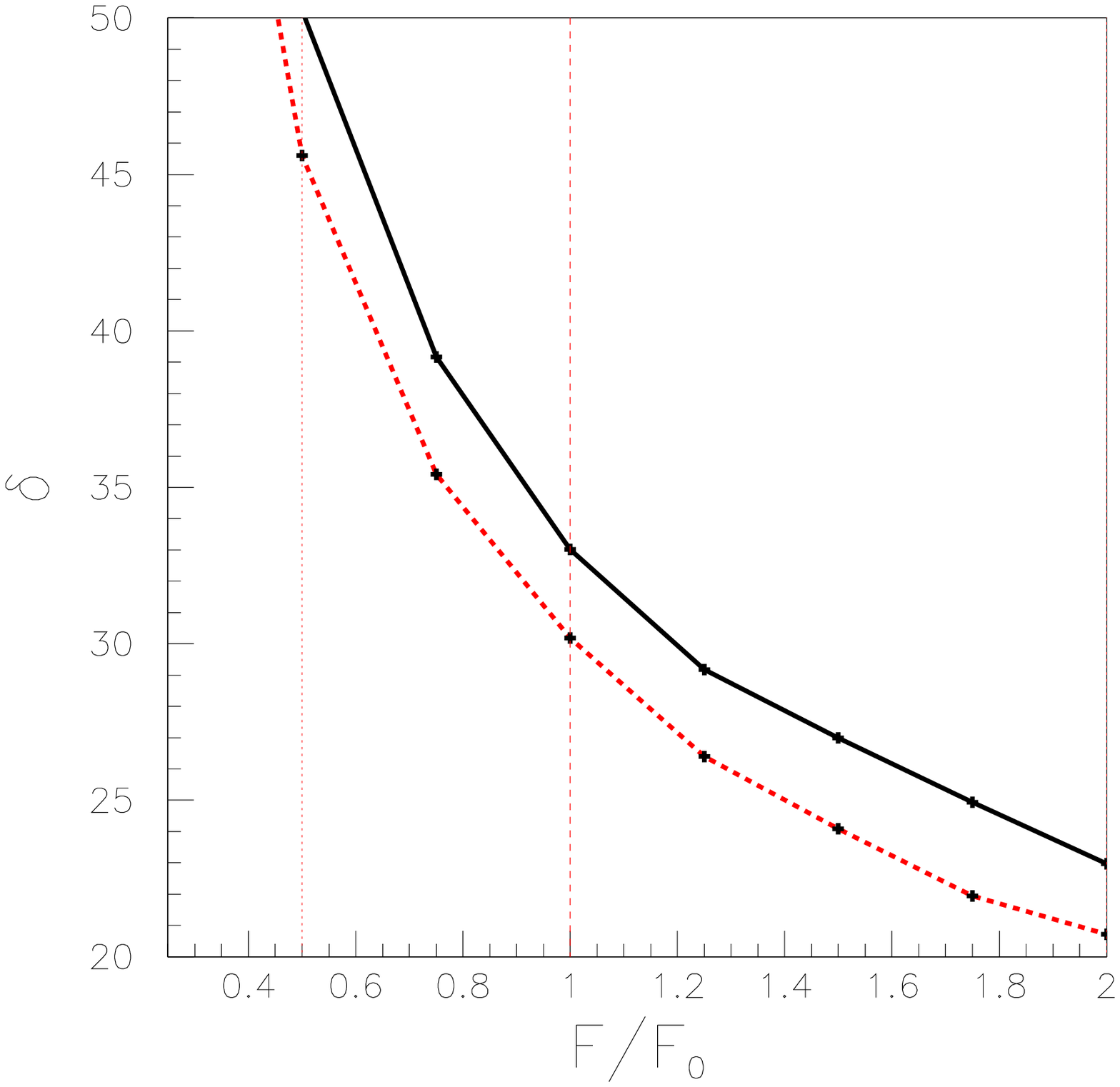}
\caption{Left plot: minimum $\theta_{13}$ where CP violation can be
established at 99\% C.L. for $\delta=90^\circ$ as a function of the
flux (1 corresponds to $F_0$). Right plot: minimum $\delta$ that can
be distinguished from zero, at 99\% C.L., as a function of the
neutrino flux for $\theta_{13}=3^\circ$. Black (dark) line corresponds to
the $\gamma=350,350$ option, red (light) to the  $\gamma=350,580$ option.}
  \label{fluxstudy}
\end{figure*}

\begin{figure}[tbph]
\centering
\includegraphics[width=75mm]{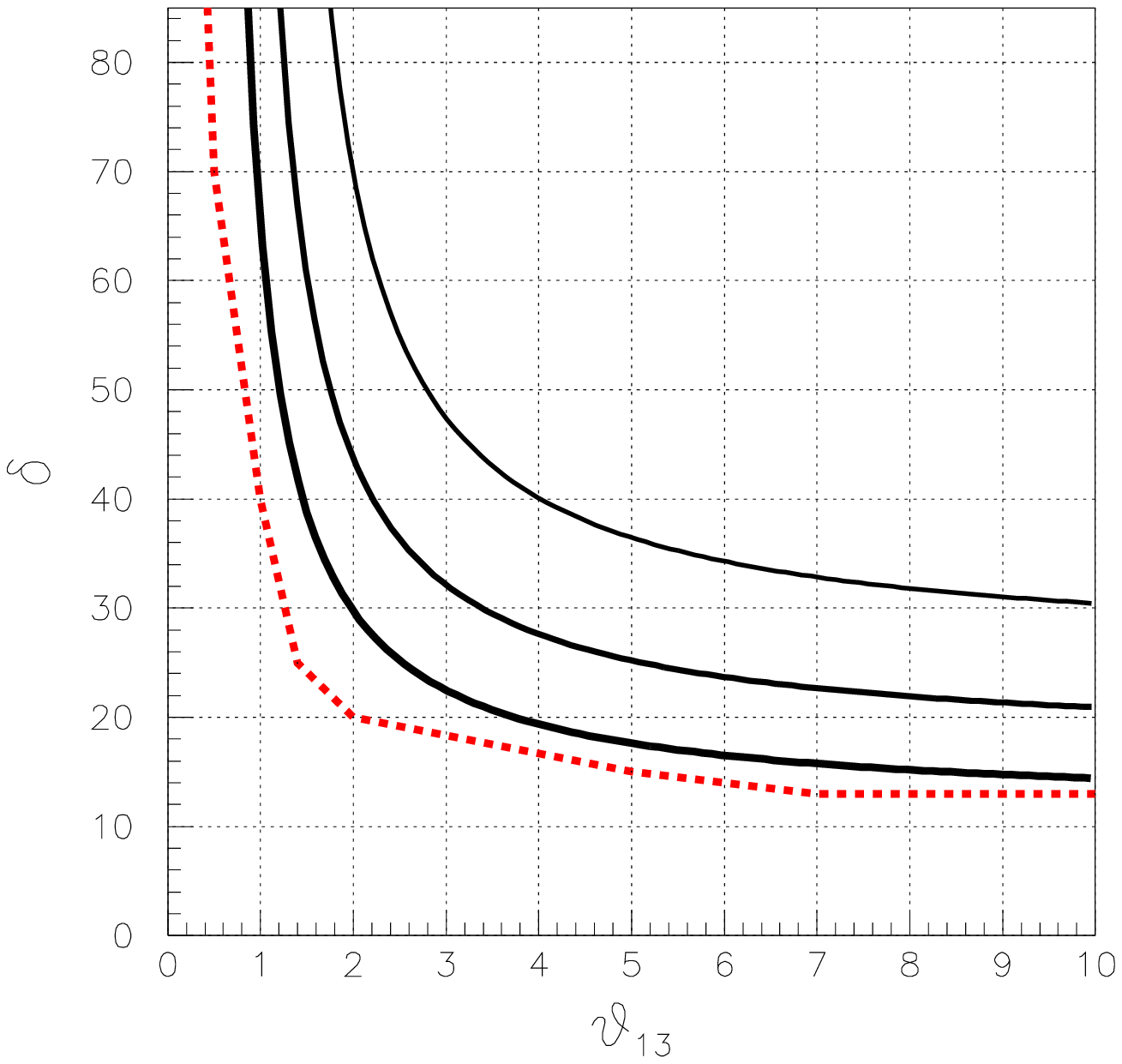}
\includegraphics[width=75mm]{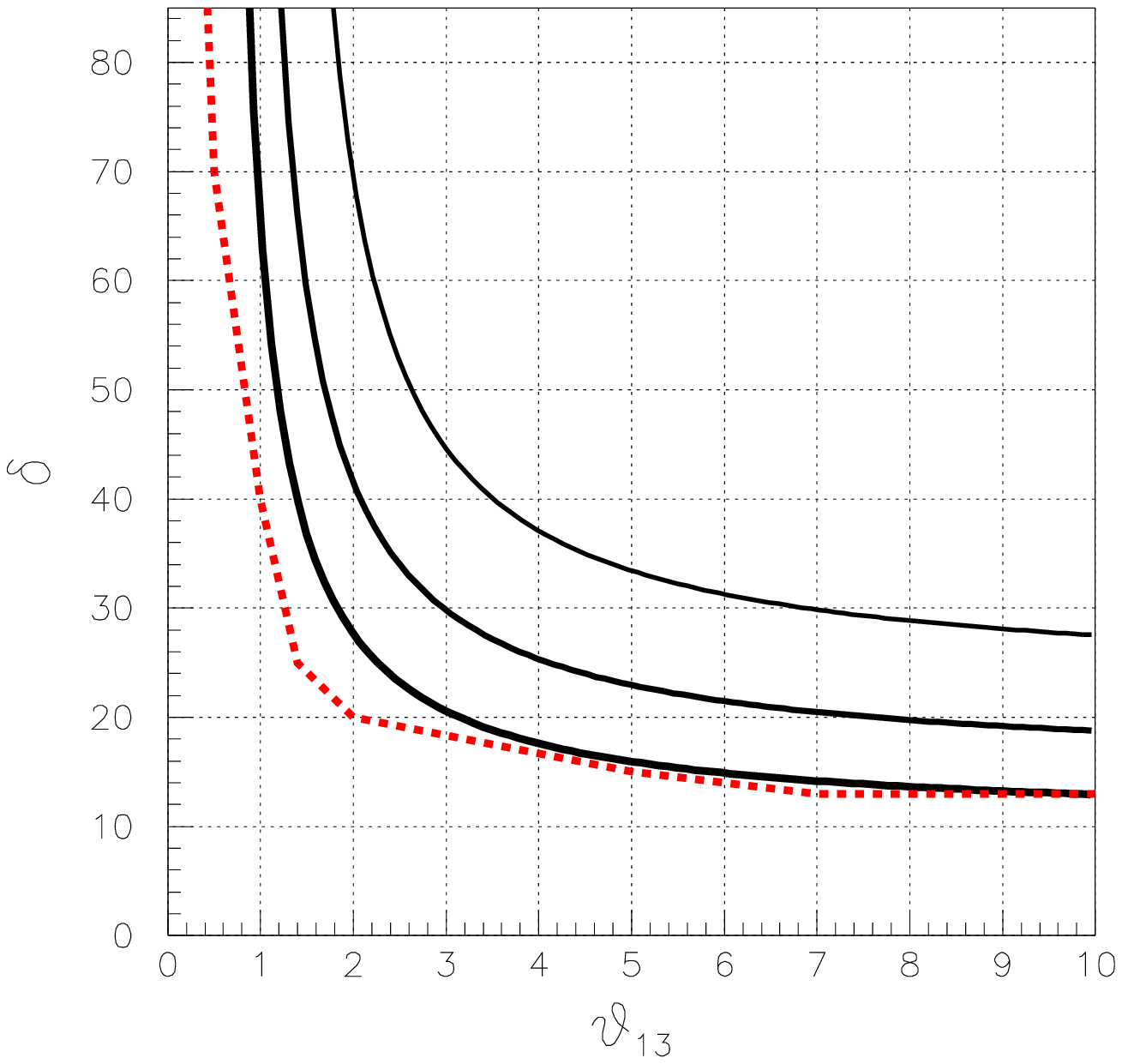}
\caption{ $\delta_{CP}$ discovery potential at 99\% C.L. as a function
of $\theta_{13}$ for the $\gamma=350,350$ (left plot) and
$\gamma=350,580$ (right plot) option.  The different solid lines
corresponds to different fluxes. From left to right: $2{\times}F_0$,
$F_0$ and $F_0/2$. The dashed line show the discovery
potential for the baseline Beta Beam option as computed in
Ref.~\cite{mauronufact}.  }
  \label{deltadisc}
\end{figure}

The main task of the second generation of accelerator experiments
(beyond T2K or NO$\nu$A) is the search for an additional source of CP
violation in the universe coming from three-family leptonic mixing. As
noted above, significant constraints on the size of $\delta$ can be
put only if $\theta_{13}$ is not highly suppressed. The next
generation of long baseline and reactor experiments is designed to
explore $\theta_{13}$ regions down to $\sim 3^\circ$ and a positive
result will likely trigger the construction of ``Phase II'' facilities
as the one discussed in this paper. Therefore, it is of paramount
importance to test down to what $\delta$ value CP violation can be
established for {\it any} value of $\theta_{13}$ that can be accessed
by T2K or NO$\nu$A. For the present facility, this is shown in
Fig.\ref{fluxstudy} as a function of the flux ($F_0$ corresponds to
nominal fluxes for $\helio$ and $\neon$). In particular, the
horizontal bands indicate the regions excluded at 90\%~CL by
T2K~\footnote{Exclusion limits depends significantly on the true
(unknown) value of $\delta$, especially if no antineutrino run is
foreseen~\cite{Migliozzi:2003pw}. In the plot, limits at
$\delta=90^\circ,0^\circ,-90^\circ$ are indicated.}. Clearly, even at
fluxes significantly smaller than $F_0$, maximal ($\delta=90^\circ$)
CP violation can be established. For values of $\theta_{13}=3^\circ$
(where T2K has reasonable chance to report solid evidence of $\nu_e$
appearance for a wide range of $\delta$ values) this Beta Beam
facility operated at nominal fluxes can establish CP violation (at
99\% CL) down to $\delta\sim30^\circ$ (see Fig.\ref{fluxstudy} right
plot).  The minimum $\delta_{CP}$ that can give evidence of CPV at
99\% C.L., as a function of $\theta_{13}$, is shown in
Fig.~\ref{deltadisc} for various fluxes.  To ease comparison with
current literature, the discovery potential of the SPS-based Beta Beam
with a Mton-size water Cherenkov detector (``baseline option'') is
also reported\footnote{For a comparison with water Cherenkov detectors
operated at high $\gamma$ Beta Beams
see~\cite{Burguet-Castell:2003vv}.}~\cite{mauronufact,
Campagne:2006yx}.

\begin{figure}[tbph]
\centering
\includegraphics[width=75mm]{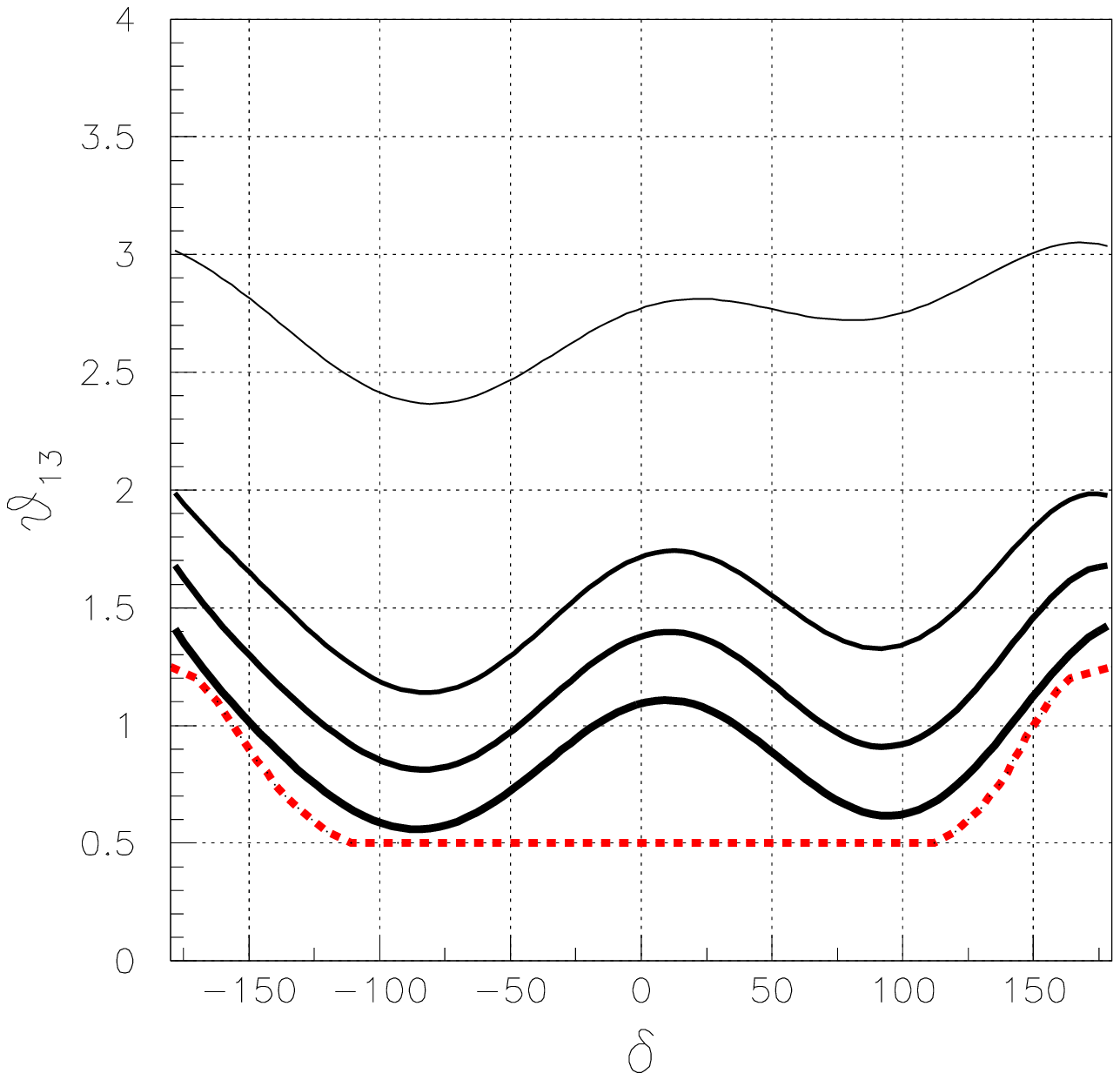}
\includegraphics[width=75mm]{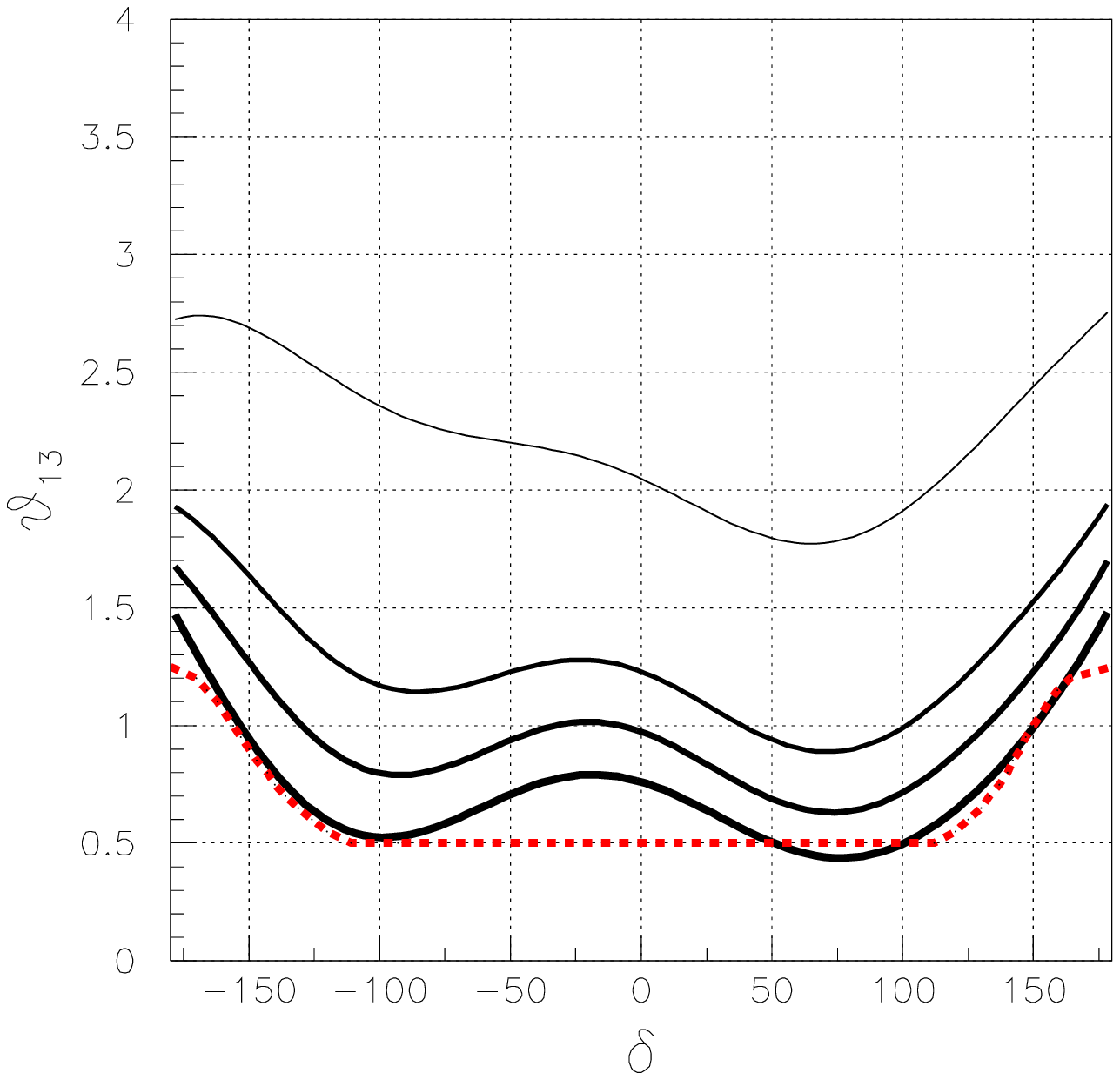}
\caption{$\theta_{13}$ limits at 90\% C.L. as a function of $\delta$
for the $\gamma=350,350$ (left plot) and $\gamma=350,580$ (right plot)
option.  The different solid lines corresponds to different
fluxes. From down to top: $2{\times}F_0$, $F_0$, $F_0/2$ and
$F_0/10$. The dashed line show the limits for the baseline scenario as
computed in Ref.~\cite{mauronufact}.}
  \label{exclplot}
\end{figure}

In case of null result\footnote{This implies the possibility of a
second generation facility to be built independently from the outcome
of T2K, NO$\nu$A and the novel reactor experiments.} additional
constraints can be put in the $\theta_{13}-\delta$ parameter
phase. They are shown in Fig.\ref{exclplot} together with the limits
from the baseline Beta Beam option.

\subsection{Neutrino hierarchy}

\begin{figure*}[tbhp]
\centering
\psfig{file=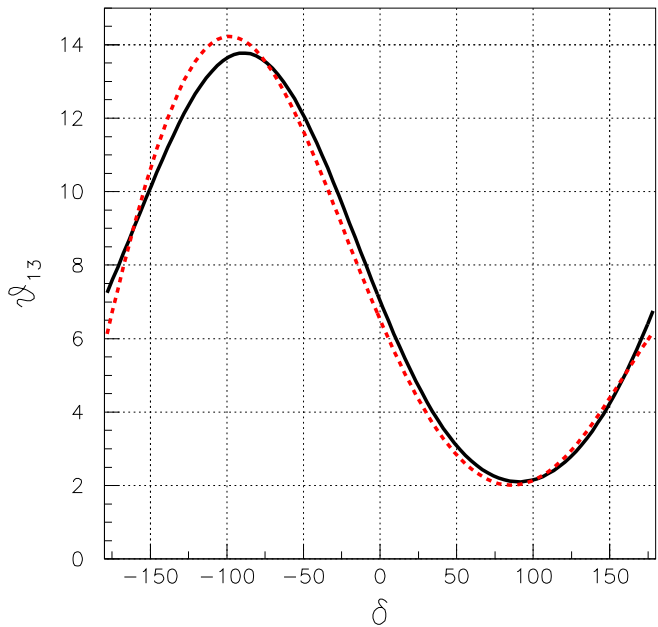,width=10cm}
\caption{Region of the parameter space where it is possible to
distinguish at 99\%C.L. the (true) hypothesis $\Delta m^2_{23}>0$ from
the $\Delta m^2_{23}<0$. Black (dark) line corresponds to the
$\gamma=350,350$ option, red (light) to the $\gamma=350,580$ option. }
  \label{signdm2}
\end{figure*}

As noted above, matter effects perturb the transition probabilities
and a simultaneous fit of the energy distributions for neutrinos and
antineutrinos allows to fix the neutrino hierarchy (sign of $\Delta
m^2_{23}$) in large areas of the $\theta_{13}-\delta$
plane. Fig.\ref{signdm2} shows the $\theta_{13}-\delta$ region where
the sign of $\Delta m^2_{23}$ can be established to be positive at 99\% C.L.
(normal hierarchy; the plot for inverted hierarchy is almost symmetric with
respect to the $\delta=0$ axis). As already discussed, for $\Delta
m^2_{23}>0$ ($\Delta m^2_{23}<0$) the sensitivity mainly resides in
regions of positive (negative) $\delta$.

\section{Conclusions}
\label{sec:conclusions}

It is an established fact that the physics case for a CERN-based Beta
Beam is limited by the smallness of the outgoing neutrino energy.  The
CERN design is based on the present LHC injection complex and it is
limited by the rigidity of the SPS and the intensity that can be
handled at PS. On the other hand, removal of these constraints is
highly beneficial to the LHC itself. In particular, the construction of
a fast cycling 1~TeV machine (``Super-SPS'') as the one proposed in
the context of the ``Phase II'' lumi upgrade of the LHC can improve
substantially the physics reach of the European Beta Beam. In this
paper, we discussed in detail the synergies between the LHC machine
upgrades and the Beta Beam technology.  At the energies that can be
reached by the Super-SPS, we demonstrated that dense detectors (iron
calorimeters) can already be used. Even at moderate masses (40 kton)
as the one imposed by the use of existing underground halls at Gran
Sasso, the CP reach is very large for any value of $\theta_{13}$ that
would provide evidence of $\nu_e$ appearance at T2K or NO$\nu$A. Moreover,
exploitation of matter effects (impossible with SPS-based options) add
sensitivity to the neutrino hierarchy. Therefore, the Beta Beam
represents a very relevant enhancement of the case for a fast cycling
1~TeV injector at CERN.

\section*{Acknowledgments}
We wish to express our gratitude to F.~Ferroni, M.~Lindroos, M.~Mezzetto
F.~Ronga and W.~Scandale for many useful discussions.


\end{document}